\documentclass[prb,twocolumn,amsmath,amssymb,floatfix]{revtex4}
\usepackage{graphicx}

\newcommand{\be}{\begin{eqnarray}}
\newcommand{\ee}{\end{eqnarray}}
\newcommand{\bfk}{{\bf k}}
\newcommand{\bfp}{{\bf p}}
\newcommand{\bfq}{{\bf q}}

\newcommand{\wbe}{\begin{widetext}}
\newcommand{\wee}{\end{widetext}}

\newcommand{\hf}{\hat{f}}
\newcommand{\hb}{\hat{b}}
\newcommand{\hrho}{\hat{\rho}}
\newcommand{\hbeta}{\hat{\beta}}

\begin{document}
\draft

\title{Strong coupling theory for the superfluidity of 
Bose-Fermi mixtures}

\author{Daw-Wei Wang}

\address{Department of Physics, National Tsing-Hua University,
Hsinchu, Taiwan 300, Republic of China}

\date{\today}

\begin{abstract}
We develop a strong coupling theory for the superfluidity of 
fermion pairing phase in a Bose-Fermi mixture.
Dynamical screening, self-energy renormalization, and pairing
gap function are included self-consistently within the adiabatic limit (i.e.
the phonon velocity is much smaller than the Fermi velocity).
An analytical solution for the transition temperature ($T_c$) 
is derived within reasonable approximations.
Using typical parameters of a $^{40}$K-$^{87}$Rb mixture, we find that
the calculated $T_c$ is several times larger than that obtained in 
the weak coupling theory,
and can be up to several percents of the Fermi temperature.
\end{abstract}


\maketitle
Recently superfluidity of fermion pairing phase
has become an extensively studied subject in the field of ultracold atoms
[\onlinecite{direct_pairing,bec_book,viverit_swave,matera,BEC_BCS,d-wave,BFM_other}].
The attractive interaction for two fermions to form a Cooper
pair can be provided either by the direct $s$-wave 
scattering between different spin states 
[\onlinecite{direct_pairing,bec_book,BEC_BCS}],
or by the effective interaction mediated
by the condensate phonons in a Bose-Fermi mixture
(BFM) [\onlinecite{viverit_swave,matera}].
The former systems attract a lot of attention on the
BCS-BEC crossover near the Feshbach resonance regime [\onlinecite{BEC_BCS}],
while the latter systems, which can mimic some traditional electron-phonon systems 
in solid state physics, are expected to have many interesting many-body
phases in different parameter regimes
[\onlinecite{d-wave,BFM_other,Collapse_exp}].

To observe the predicted fermion pairing phase or other many-body
states in a BFM, experimentalists usually have to
increase the boson-fermion interaction strength 
via a heteronuclear Feshbach resonance
[\onlinecite{BFM_other,BFM_exp1}].
However, by doing so the system will be driven into a
strong coupling regime,
where the direct application of the Bardeen-Cooper-Schrieffer (BCS) 
theory for the superfluidity of fermion pairing phase
becomes unjustified and not reliable. For example, the strong fermion-boson
interaction may renormalize the effective mass of fermions and hence 
change its density of states near the Fermi surface. Besides,
the fermionic quasi-particle fluctuations at finite temperature can also 
dynamically screen the effective interaction induced by condensate phonons.
None of the these effects have been included in the weak 
coupling theory developed in the literature 
[\onlinecite{viverit_swave,matera,d-wave}].
Moreover, in most of the weak coupling theories, 
one usually applys an instantaneous approximation (i.e. assuming
the phonon velocity, $c_b$, is much larger than the Fermi velocity, $v_F$).
Such approximation is, however, very unrealistic in typical
systems like $^{40}$K-$^{87}$Rb or $^6$Li-$^{23}$Na mixtures,
which are usually in the adiabatic regime ($c_b/v_F\sim 0.05-0.5<1$)
due to the weak interaction between bosonic atoms [\onlinecite{matera}].
Therefore in order to have a reasonable comparison 
between the theoretical results and the experimental measurement,
it is necessary and important 
to develop a full strong coupling theory, including the adiabaticity
of condensate phonons as well as the strong correlation effects
of a BFM self-consistently.

In this Letter, we develop a strong coupling theory for the superfluidity
of ($s$-wave) fermion pairing phase in a BFM within the adiabatic 
regime by generalizing
the celebrated Migdal-Eliashberg equations used in the conventional
superconductors [\onlinecite{migdal,RMP}]. By strong coupling, we 
mean to treat the fermion self-energy, dynamical screening of 
the effective interaction,
and the pairing gap function self-consistently. Hence the obtained
results are reliable even when the fermion-boson interaction is not weak.
Vertex diagrams are safely neglected by applying Migdal's theorem 
in the adiabatic regime of phonons
[\onlinecite{migdal,RMP}].
Using a single mode approximation we further 
derive an analytical solution for the 
superfluidity transition temperatures and compare them with the known 
weak coupling BCS results in different parameter regime.
In a typical $^{40}$K-$^{87}$Rb mixture,
we find that the obtained $T_c$ is several times larger than 
the weak coupling results, and can be up to 
a few percents of the Fermi energy before reaching phase instability.
Effects due to strong boson-fermion correlations are also 
critically discussed.

We consider a three dimensional BFM composed of
spin polarized bosons and fermions
in two equally occupied hyperfine spin states. 
For simplicity, we neglect the inhomogeneous magnetic trap potential 
so that the total Hamiltonian can be written to be [\onlinecite{d-wave}]
\begin{eqnarray}
H &=&\sum_{\bfk} \left[
\bar{\epsilon}^b_\bfk \hb_\bfk^{\dagger}\hb_\bfk
+ \bar{\epsilon}^f_{\bfk,\uparrow} 
\hf_{\bfk,\uparrow}^{\dagger}\hf_{\bfk,\uparrow}
+ \bar{\epsilon}^f_{\bfk,\downarrow} \hf_{\bfk,\downarrow}^{\dagger}
\hf_{\bfk,\downarrow}
\right]
\\
&&+\frac{1}{\Omega}\sum_\bfk\left[
\frac{U_{bb}}{2}\hrho^b_\bfk \hrho^b_{-\bfk}
+U_{bf}\hrho^b_{\bfk} \hrho^f_{-\bfk}
+U_{ff}\hrho^f_{\bfk,\uparrow} \hrho^f_{-\bfk,\downarrow}\right]
\nonumber
\label{H_tot_spinful}
\end{eqnarray}
where $\hb_\bfk$ and $\hf_{\bfk,s}$ are the field operators 
for bosonic and fermionic atoms of momentum $\bfk$ 
and isospin $s=\uparrow,\downarrow$.
$\hrho^b_\bfk=\sum_\bfp \hb^\dagger_{\bfp+\bfk}\hb^{}_\bfp$
is the boson density operator, and 
$\bar{\epsilon}^{b}_\bfk=\bfk^2/2m_b-\mu_b$ 
is the bosonic kinetic energy with $m_{b}$ being the atom mass
and $\mu_b$ being the chemical potential. 
Similar notations also apply to fermions with super/subscript
$f$, and $\hrho^f_\bfk\equiv\hrho^f_{\bfk,\uparrow}
+\hrho^f_{\bfk,\downarrow}$.
$U_{bb}=\frac{4\pi a_{bb}}{m_b}$, $U_{ff}=\frac{4\pi a_{ff}}{m_f}$,
and $U_{bf}=\frac{2\pi a_{bf}}{m_r}$, are respectively
the boson-boson, fermion-fermion and boson-fermion pseudo-potential 
strength, with $a_{ij}$ being the associate $s-$wave scattering length.
$m_r$ is the reduced mass and $\Omega$ is the system volume.

\begin{figure}
\includegraphics[width=8cm]{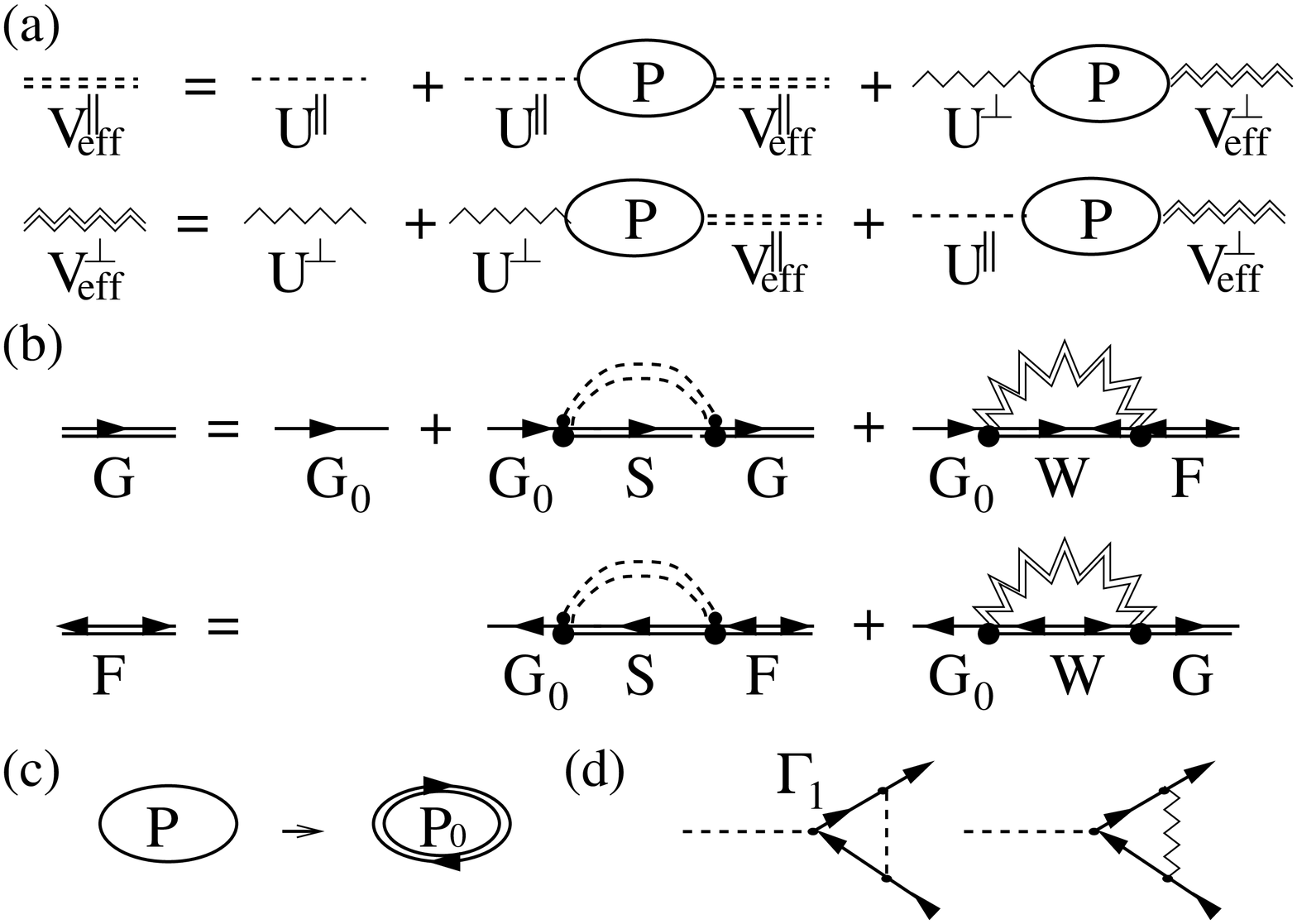}
\caption{
Feynman diagrams studied in the strong coupling theory of a BFM.
(a) Full (dressed) effective interaction (see the text); (b)
Migdal-Eliashberg equations, where $S$ and $W$ are the self-energies for
normal and anomalous Green's functions respectively; 
(c) Polarizability, $P$, is
replaced by bubbles without vertex correction;
(d) Vertex correction neglected in the strong coupling theory due
to Migdal's theorem.
}
\label{V_eff_com}
\end{figure}
\begin{figure}
\includegraphics[width=7cm]{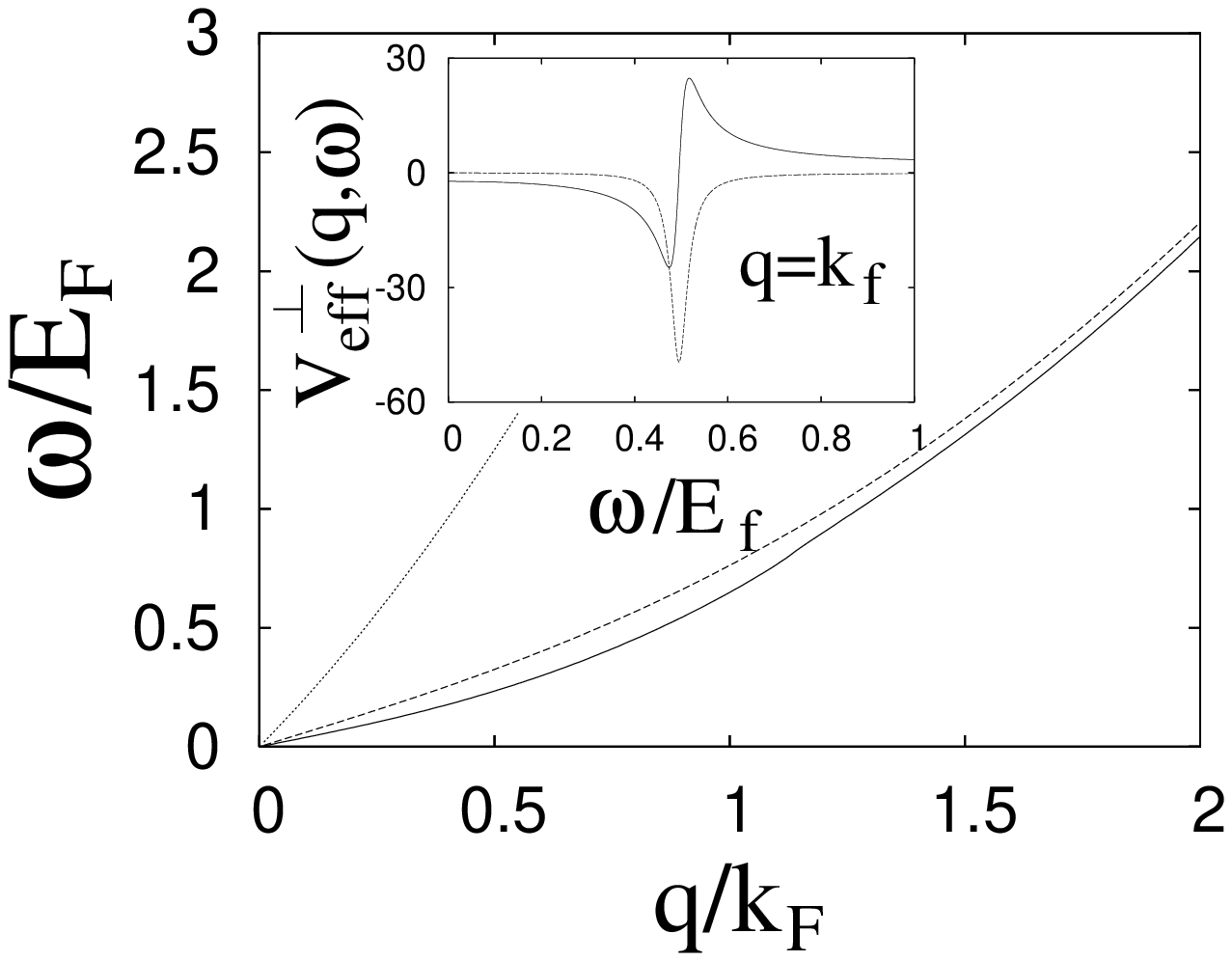}
\caption{
Phonon dispersions including (solid line) and not-including (dashed line) the 
dynamical screening effects. 
Below the dotted line is regime of quasi-particle(fermion) excitations,
where the Landau damping broadens the phonon spectral function
as shown in the inset.
Inset: the real(solid line) and imaginary(dashed line) parts of the full
effective interaction, $V^\perp_{\rm eff}(q,\omega)$, at $q=k_F$. 
Parameters are in Ref. [\onlinecite{parameter}].
}
\label{mode}
\end{figure}

Throughout this Letter we assume that the temperature is so low that all
bosons are condensed at zero momentum state.
The condensate excitations (phonons) are of Bogoliubov
type dispersion: $\omega_\bfk^0 = c_b |\bfk|\sqrt{1+|\bfk|^2\xi_b^2}$, where
$c_b\equiv\sqrt{n_bU_{bb}/m_b}$ and $\xi_b\equiv\sqrt{1/4m_bn_bU_{bb}}$ 
are respectively the phonon velocity and the healing length of the
condensate with density $n_b$. As a result, one can effectively
describe the BFM system by the following fermion-phonon type Hamiltonian
[\onlinecite{viverit_swave,matera,d-wave}]:
$
H=\sum_{\bfk,s}\bar{\epsilon}^f_\bfk \hf^\dagger_{\bfk,s} \hf_{\bfk}^{}
+\sum_\bfk\left[\omega^0_\bfk\beta_\bfk^\dagger\beta_\bfk^{}
+g_\bfk(\hbeta_\bfk+\hbeta_{-\bfk}^\dagger)\hrho_{-\bfk}^f\right]
$
where $\hbeta_\bfk$ is the phonon operator and
$g_\bfk = U_{bf}\sqrt{n_b\epsilon^b_\bfk/\omega^0_\bfk}$ measures the
fermion-phonon coupling strength. Integrating out the phonon field, 
one can obtain a retarded phonon-induced interaction between fermions:
$V_{\rm ph}(\bfk,\omega)
=\frac{-2{g}_\bfk^2\omega^0_\bfk}{(\omega^0_\bfk)^2-(\omega+i\delta)^2}$. 
When the phonon velocity is smaller than the Fermi velocity
as considered in this Letter, the dynamical screening due to
fermion density fluctuations has to be included by calculating the
Dyson's equations shown in Fig. \ref{V_eff_com}(a), where
$U^{\|}=V_{\rm ph}$ and $U^{\perp}=V_{\rm ph}+U_{ff}$ are the bare
effective interaction between fermions
in the spin parallel/perpendicular channels. The full dressed effective
interaction can then be easily derived to be [\onlinecite{RMP}]:
\be
V_{\rm eff}^{\|,\perp} &=& 
\frac{(1-U^{\|,\perp}P)U^{\|,\perp}+U^{\perp}PU^{\perp}}
{(1-U^{\|}P)^2-(PU^{\perp})^2}
\label{V_||}
\ee
where $P(\bfk,\omega)$ is the polarizability due to fermionic quasi-particle
and Cooper pair fluctuations. In Fig. \ref{mode} we show a numerical 
result of $V^\perp_{\rm eff}(q,\omega)$ and the screened phonon
dispersion (obtained by tracking the peak position of 
${\rm Im}V^\perp_{\rm eff}(q,\omega)$) at $T=T_c$ for a typical
$^{40}$K-$^{87}$Rb system [\onlinecite{parameter}]. 
As we will see later, although
the screened phonon dispersion is close to the bare 
(unscreened) results, their spectral weights can still be
quiet different due to the fluctuations near the Fermi surface.

Now the single particle
Green's function, $G(\bfp,t)=-i\langle \hat{T} f^{}_{\bfp,\sigma}(t)
f^\dagger_{\bfp,\sigma}(0)\rangle$, and the anomalous Green's function
$F(\bfp,t)=-i\langle \hat{T} f^{}_{-\bfp,\downarrow}(t)
f^{}_{\bfp,\uparrow}(0)\rangle$ (where $\hat{T}$ is the time-ordering operator)
can be calculated within a self-consistent Hartree-Fock approximation
as shown in Fig. \ref{V_eff_com}(b).
Such meanfield treatment is justified in strong 
fermion-phonon interaction regime because the vertex correction
is proportional to the ratio of the phonon velocity to the Fermi velocity 
(known as Migdal's theorem [\onlinecite{migdal}])
and hence its leading order correction to the bare vertex 
(Fig. \ref{V_eff_com}(d))
can be rather small ($\Gamma_1<0.1$, see Ref. [\onlinecite{parameter}]) 
in typical $^{40}$K-$^{87}$Rb or $^{6}$Li-$^{23}$Na mixtures.
By the same reason, we can also approximate the full fermion
polarizability, $P$, by a bubble diagram, 
$P_0(\bfk,\omega)=\frac{2}{i\Omega}\sum_\bfp\int\frac{d\nu}{2\pi}
G(\bfp,\nu)G(\bfp+\bfk,\nu+\omega)$,
without vertex lines inside the bubble diagram (Fig. \ref{V_eff_com}(c)).
The contribution coming from pair fluctuations 
can be neglected since here we are only interested in the 
calculation of $T_c$.

To formulate the Migdal-Eliashberg equations 
(MEE, see Fig. \ref{V_eff_com}(b)) for a BFM system,
we note that the phonon excitation energy, $\omega_\bfk^0$, 
is unbounded in the high energy limit, and hence
the involved momentum integration
cannot be restricted to the vincinity of the Fermi surface, as done in the original
theory for the conventional superconductors [\onlinecite{RMP}].
We therefore define the following energy dependent interaction strength:
\be
\widetilde{\alpha^2F}_{\|,\perp}(\nu,E)&\equiv&
\int_{FS} \frac{d^2\bfp
B_{\|,\perp}(\bfp-\sqrt{2m_fE}\hat{z},\nu)}{v_F(2\pi)^3N(E_F)/N(E)},
\label{a2F}
\ee
where $B_{\|,\perp}(\bfq,\nu)\equiv -\pi^{-1}{\rm Im}
V^{\|,\perp}_{\rm eff}(\bfq,\nu)$ is the spectral weight of the dressed
interaction; $\int_{FS}$ denotes the integration over Fermi surface, 
where the gap function is measured.
$N(E)=m_f\sqrt{2m_f E}/2\pi^2$ is th 3D density of states for $E>0$.
Note that $\widetilde{\alpha^2F}_{\|,\perp}(\nu,E)$ drops to zero rapidly
for $E\to\infty$ due to the sharp spectral function, $B_{\|,\perp}$,
near the phonon excitation energy (see Inset of Fig. \ref{mode}).
Following the same argument as Eliashberg [\onlinecite{RMP}], we can 
neglect the momentum dependence of self-energies, $S$ and $W$, and then solve 
Fig. \ref{V_eff_com}(b) by introducing the gap function, 
$\Delta(\omega)\equiv W(\omega)/Z(\omega)$, and a renormalization coefficient,
$Z(\omega)\equiv 1-(S(\omega)-S(-\omega))/2\omega$ [\onlinecite{RMP}].
$Z(\omega\to 0)^{-1}\leq 1$ measures 
the single particle spectral weight at Fermi surface, and 
is equal to one for noninteracting fermions.
Details of the derivation for a BFM system will be
presented elsewhere [\onlinecite{unpublished}].
The obtained MEE are the following:
\wbe
\be
\omega(1-Z(\omega))&=&-
\int_0^\infty d\nu\int_{-\infty}^\infty d\epsilon
\,\widetilde{\alpha^2F}_\|(\nu,\mu+\epsilon \tilde{Z}(\epsilon))
{\rm Re}\left[\frac{|\epsilon|}{\sqrt{\epsilon^2-\Delta(\epsilon)^2}}\right]
\left[\frac{n_F(\epsilon)+n_B(\nu)}
{\epsilon-\omega-\nu-i\delta}
+\frac{1-n_F(\epsilon)+n_B(\nu)}{\epsilon-\omega+\nu-i\delta}\right]
\label{eq:K-0}
\\
\Delta(\omega)Z(\omega) &=&
U_{ff}\int_{-\infty}^\infty d\epsilon\, n_F(\epsilon)
{\rm Re}\left[\frac{\Delta(\epsilon){\rm sgn}(\epsilon)}
{\sqrt{\epsilon^2-\Delta(\epsilon)^2}}\right]
N\left(\mu+\epsilon \tilde{Z}(\epsilon)\right)
\nonumber\\
&&+\int_0^\infty d\nu\int_{-\infty}^\infty d\epsilon
\,\widetilde{\alpha^2F}_\perp(\nu,\mu+\epsilon \tilde{Z}(\epsilon))
{\rm Re}\left[\frac{\Delta(\epsilon){\rm sgn}(\epsilon)}
{\sqrt{\epsilon^2-\Delta(\epsilon)^2}}\right]
\left[\frac{n_F(\epsilon)+n_B(\nu)}
{\epsilon-\omega-\nu-i\delta}
+\frac{1-n_F(\epsilon)+n_B(\nu)}{\epsilon-\omega+\nu-i\delta}\right]
\label{eq:K+0}
\ee
\wee
where $n_{F/B}(\epsilon)=(e^{\epsilon/T}\pm 1)^{-1}$ is the 
Fermi/Bose distribution function.
$\tilde{Z}(\epsilon)$ is the real part of
$Z(\epsilon)\sqrt{1-(\Delta(\epsilon)/\epsilon)^2}$.
Note that the instantaneous interaction, $U_{ff}$,
contributes to the self-energy, $S$, via a constant shift
only [\onlinecite{RMP}] and therefore does not appear in the 
r.h.s. of Eq. (\ref{eq:K-0}).
Furthermore, the ultraviolet divergence caused by the 
pseudo-potential $U_{ff}$ 
has also been removed in Eq. (\ref{eq:K+0})
by eliminating the bare Lippmann-Schwinger equations 
{\it in vacuo} as done in the weak coupling theory 
[\onlinecite{bec_book,viverit_swave}]. Therefore the MEE shown above
have included both the non-retarded direct interaction 
and the retarded induced interaction 
on equal footing.

To calculate the $T_c$ of Cooper pairs 
we first linearize the above equations by taking
$\Delta(\omega)\to 0$ and approximate $Z(\omega)$
by $Z_0=Z(\omega\to 0)$.
Assuming $T_c$ is much smaller than Fermi energy and typical phonon energy, 
we can simplify Eqs. (\ref{eq:K-0})-(\ref{eq:K+0}) to be
\be
1-Z_0&=&-2\int_0^\infty d\nu \int_0^\infty d\epsilon\,
\frac{\widetilde{\alpha^2F}_\|(\nu,\mu+Z_0\epsilon)}
{(\epsilon+\nu)^2}
\label{eq1}
\\
Z_0 &=& U_{ff}\int_{-\infty}^\infty \frac{d\epsilon}{\epsilon}\, 
n_F(\epsilon) N\left(\mu+\epsilon Z_0\right)
\nonumber
\ee
\be
&&+\int_0^\infty d\nu\int_{-\infty}^\infty 
\frac{n_F(\epsilon)d\epsilon}{\epsilon(\epsilon-\nu)}
\widetilde{\alpha^2F}_\perp(\nu,\mu+Z_0\epsilon).
\label{eq2}
\ee
To solve $T_c$ and $Z_0$ analytically we further apply 
the single mode approximation, 
$
{\rm Im}V_{\rm eff}^{\|,\perp}
(q,\nu)\sim -\pi V_{\rm im}^{\|,\perp}(q)\delta(\nu-\omega_q),
$
where $V^{\|,\perp}_{\rm im}(q)\equiv 
-\frac{1}{\pi}\int_0^\infty d\nu\, {\rm Im}V^{\|,\perp}_{\rm eff}(q,\nu)$
is the total spectral weight.
The phonon energy, $\omega_q$, is obtained from 
the peak position of ${\rm Im}V^{\|,\perp}_{\rm eff}(q,\nu)$.
Within such approximation $\widetilde{\alpha^2F}_{\|,\perp}$ of Eq. (\ref{a2F})
can be greatly simplified and $T_c$ of Eq. (\ref{eq2}) can be solved to be
\be
\frac{T_c}{E_F}&=&\left(\frac{8\gamma}{\pi e^2 Z_0}\right)
\exp\left[\frac{1+\lambda_\|(Z_0)+\lambda_{\log}(D)}{\lambda_0
-\lambda_\perp(Z_0)}\right],
\label{T_c}
\ee
where $\gamma\sim 1.781$.
$\lambda_0\equiv N(E_F)U_{ff}$ and $\lambda_\perp(Z_0)\equiv
2\int_0^\infty\frac{d\nu}{\nu}
\widetilde{\alpha^2 F}_\perp(\nu,0)$ are respectively the 
direct and induced coupling strengths for Cooper pairs.
$D\equiv 4E_F/e^2 Z_0$ is an energy scale of Fermi energy. 
$Z_0$ is then determined by solving
$Z_0=1+\lambda_\|(Z_0)$ self-consistently since $Z_0$ appears in the 
effective interaction via the polarizability, $P$. The two coupling
strengths, $\lambda_\|(Z_0)$ and $\lambda_{log}(D)$,
are given by (here $\nu_c\equiv\omega_{2 k_F}$)
\wbe
\be
\lambda_\|(Z_0)&=&\int_0^{\nu_c} d\nu A^\|(\nu)\left[
\frac{2}{\nu}-\frac{1}{\nu+|\epsilon_-(\nu)|}-\frac{1}
{\nu+\epsilon_+(\nu)}\right]
-\int_{\nu_c}^\infty d\nu A^\|(\nu)\left[\frac{1}{\nu+\epsilon_-(\nu)}
-\frac{1}{\nu+\epsilon_+(\nu)}\right]
\\
\lambda_{\log}(D)&=&
\int_0^{\nu_c}\frac{d\nu A^\perp(\nu)}{\nu}
\log\left(\frac{\nu^2|\epsilon_-|\epsilon_+}
{D^2(\nu+|\epsilon_-|)(\nu+\epsilon_+)}\right)
+\int_{\nu_c}^\infty\frac{d\nu A^\perp(\nu)}{\nu}
\log\left(\frac{\epsilon_+(\nu+\epsilon_-)}{\epsilon_-(\nu+\epsilon_+)}\right),
\ee
\wee
where $A^{\|,\perp}(\nu)\equiv \frac{N(E_F)}{2k_F^2}$
$\left|\frac{\partial\omega_q(q_\ast)}{\partial q}\right|^{-1}
q_\ast V_{\rm im}^{\|,\perp}(q_\ast)$ with $q_\ast(\nu)$ given by
$\omega_{q_\ast(\nu)}=\nu$, and
${\epsilon}_\pm(\nu)\equiv q_\ast^2/2m_f\pm q_\ast k_F/m_f$.

\begin{figure}
\includegraphics[width=9cm]{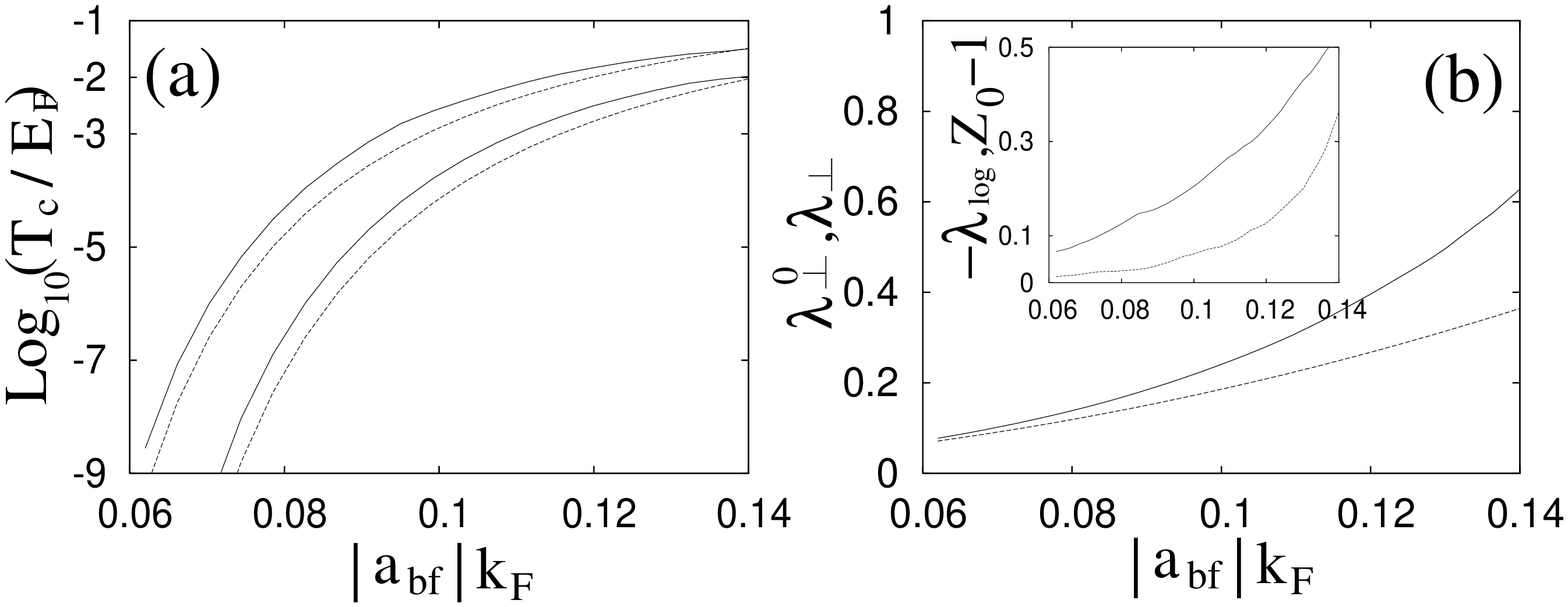}
\caption{(a) Superfluid temperature as a function of 
boson-fermion interaction strength,
$|a_{bf}|k_F$, for the strong coupling (solid lines) and weak coupling
(dashed lines) theories. 
Upper/lower pairs of lines are for boson density 
$n_b=100/50\times {\rm um}^{-3}$. 
Other parameters are given in Ref. [\onlinecite{parameter}].
(b) $\lambda_\perp$ and $\lambda_\perp^0$ ($\lambda_\|=Z_0-1$ 
and $-\lambda_{\rm log}$ for the inset) 
in the same calculation of (a) with $n_b=100$ um$^{-3}$.
}
\label{Tc_a_bf}
\end{figure}

It is instructive to compare the analytical 
$T_c$ in Eq. (\ref{T_c}) with the weak coupling results
[\onlinecite{viverit_swave}]: one can see that Eq. (\ref{T_c}) reproduces the 
weak coupling result [\onlinecite{viverit_swave}]: 
$T_c^0=\frac{8\gamma}{\pi e^2}
\exp\left[\frac{1}{\lambda_0-\lambda^0_\perp}\right]$
if $|\lambda_\|(Z_0)|,|\lambda_{\rm log}(D)|\ll 1$.
Here $\lambda^0_\perp=\frac{N(E_F)U_{bf}^2}{U_{bb}(2k_F\xi_b)^2}
\ln(1+(2k_F\xi_b)^2)$ is the phonon-induced coupling strength without
any dynamical screening [\onlinecite{viverit_swave}].
(Note that here we only discuss the $s$-wave pairing symmetry.
The non $s$-wave scattering strength is in general very small
and can be neglected.)
In Fig. \ref{Tc_a_bf}(a) we compare numerical results of
$T_c$ and $T_c^0$, using typical parameters of 
$^{40}$K-$^{87}$Rb systems [\onlinecite{parameter}].
We find both of them increase sub-exponentially as the interaction strength,
$|a_{bf}|k_F$, increases. $T_c$ is larger than $T_c^0$ by a factor
of 2-8 for $|a_{bf}|k_F<0.12$, and then becomes almost equal to $T_c^0$ 
for stronger
interaction. Note that $T_c/T_c^0$ is large for small $|a_{bf}|k_F$ simply
because the repulsive direct interaction, $U_{ff}>0$, sets the minimum
value of $|a_{bf}|k_F$ for pairing phase.
This ratio will decrease to one as $|a_{bf}|k_F\to 0$ 
if $U_{ff}\leq 0$. When $|a_{bf}|k_F>0.15$ the 
screened phonon velocity, $c_b$, becomes imaginary and the system
becomes unstable toward collapse.
This critical value of $|a_{bf}|k_F$ is a little higher than the results 
predicted by the weak coupling theory [\onlinecite{bec_book}] 
due to the reduction
of the density of states in the strong coupling theory.
Analyzing each quantity of Eq. (\ref{T_c})
more carefully (see Fig. \ref{Tc_a_bf}(b)), 
we find that
although the ratio of $\lambda_\perp$ to $\lambda_\perp^0$
increases as $|a_{bf}|k_F$ increases via dynamical screening,
the final ratio of $T_c$ to $T_c^0$ still becomes smaller 
due to the strong quasi-particle renormalization, $Z_0$
(see the inset), in the strong coupling regime.
Numerically solving the full strong coupling equations 
(Eqs. (\ref{eq:K-0})-(\ref{eq:K+0})) can also provide more
information about the gap function spectroscopy, 
$\Delta(\omega)$, and fermion single particle spectral function etc., 
which can be measured by using rf-spectroscopy and/or
Bragg scattering spectroscopy in the present experimental technique.

In summary, we have derived the full strong coupling theory for 
the fermion $s$-wave pairing phase in a Bose-Fermi mixture.
Our results apply 
to the limit of slow phonon velocity and 
hence are valid in the existing BFM system ($^{40}$k-$^{87}$Rb 
or $^6$Li-$^{23}$Na).
The predicted critical temperature for a typical $^{40}$K-$^{87}$Rb mixture
can be as high as a few percents of the Fermi energy, which should be 
acheivable by present experimental groups.

We thank fruitful discussion with J.H. Bao, E. Demler, M.D. Lukin, 
and C.-Y. Mou.



\begin{thebibliography}{99}
\bibitem{parameter}
The parameters used in this Letter are for a typical $^{40}$K-$^{87}$Rb system:
$^{40}$K are prepared in the hyperfine states: 
$|F,m_F\rangle=|9/2,-9/2\rangle$ 
and $|9/2,-7/2\rangle$. $^{87}$Rb are prepared in state $|1,1\rangle$. 
Fermion density, $n_f=1$ $\mu$m$^{-3}$; boson density $n_b=100$ $\mu$m$^{-3}$;
background $s$-wave scattering lengths (in unit of Bohr radius): 
$a_{ff}=174$, $a_{bb}=100$, and $a_{bf}=-330$.
Using these parameters, we have $c_b/v_F=0.6$, $k_F \xi_b=0.75$, and
the leading order vertex correction, $\Gamma_1=0.07$. 
See for example the experimental parameters 
of Ref. [\onlinecite{Collapse_exp}].
\end{thebibliography}
\end{document}